%
%
%
%
\def\unredoffs{} \def\redoffs{\voffset=-.40truein\hoffset=-.40truein}
\def\speclscape{}
%
%
%
%
\newbox\leftpage \newdimen\fullhsize \newdimen\hstitle \newdimen\hsbody
\tolerance=1000\hfuzz=2pt
\catcode`\@=11 
\def\bigans{b }
\def\answ{b }
\ifx\answ\bigans\message{(This will come out unreduced.}
\magnification=1200\unredoffs\baselineskip=16pt plus 2pt minus 1pt
\hsbody=\hsize \hstitle=\hsize 
\else\message{(This will be reduced.} \let\l@r=L
\magnification=1000\baselineskip=16pt plus 2pt minus 1pt
\vsize=7truein \redoffs
\hstitle=8truein\hsbody=4.75truein\fullhsize=10truein\hsize=\hsbody
\output={\ifnum\pageno=0 
   \shipout\vbox{\speclscape{\hsize\fullhsize\makeheadline}
     \hbox to \fullhsize{\hfill\pagebody\hfill}}\advancepageno
   \else
   \almostshipout{\leftline{\vbox{\pagebody\makefootline}}}\advancepageno
   \fi}
\def\almostshipout#1{\if L\l@r \count1=1 \message{[\the\count0.\the\count1]}
       \global\setbox\leftpage=#1 \global\let\l@r=R
  \else \count1=2
   \shipout\vbox{\speclscape{\hsize\fullhsize\makeheadline}
       \hbox to\fullhsize{\box\leftpage\hfil#1}}  \global\let\l@r=L\fi}
\fi
%
\newcount\yearltd\yearltd=\year\advance\yearltd by -1900

\def\Title#1#2{\nopagenumbers\abstractfont\hsize=\hstitle\rightline{#1}%
\vskip 1in\centerline{\titlefont #2}\abstractfont\vskip
.5in\pageno=0}
%
%

\def\draftmode{\message{ DRAFTMODE }\def\draftdate{{\rm preliminary draft:
\number\month/\number\day/\number\yearltd\ \ \hourmin}}%
\headline={\hfil\draftdate}\writelabels\baselineskip=20pt plus 2pt
minus 2pt
  {\count255=\time\divide\count255 by 60 \xdef\hourmin{\number\count255}
   \multiply\count255 by-60\advance\count255 by\time
   \xdef\hourmin{\hourmin:\ifnum\count255<10 0\fi\the\count255}}}
\def\nolabels{\def\wrlabeL##1{}\def\eqlabeL##1{}\def\reflabeL##1{}}
\def\writelabels{\def\wrlabeL##1{\leavevmode\vadjust{\rlap{\smash%
{\line{{\escapechar=` \hfill\rlap{\sevenrm\hskip.03in\string##1}}}}}}}%
\def\eqlabeL##1{{\escapechar-1\rlap{\sevenrm\hskip.05in\string##1}}}%
\def\reflabeL##1{\noexpand\llap{\noexpand\sevenrm\string\string\string##1}}}
\nolabels
%
\global\newcount\secno \global\secno=0 \global\newcount\meqno
\global\meqno=1
\def\newsec#1{\global\advance\secno by1\message{(\the\secno. #1)}
\global\subsecno=0\eqnres@t\noindent{\bf\the\secno. #1}
\writetoca{{\secsym} {#1}}\par\nobreak\medskip\nobreak}
\def\eqnres@t{\xdef\secsym{\the\secno.}\global\meqno=1\bigbreak\bigskip}
\def\sequentialequations{\def\eqnres@t{\bigbreak}}\xdef\secsym{}
\global\newcount\subsecno \global\subsecno=0
\def\subsec#1{\global\advance\subsecno by1\message{(\secsym\the\subsecno. #1)}
 \ifnum\lastpenalty>9000\else\bigbreak\fi
\noindent{\it\secsym\the\subsecno. #1}\writetoca{\string\quad
{\secsym\the\subsecno.} {#1}}\par\nobreak\medskip\nobreak}
\def\appendix#1#2{\global\meqno=1\global\subsecno=0\xdef\secsym{\hbox{#1.}}
\bigbreak\bigskip\noindent{\bf Appendix #1. #2}\message{(#1. #2)}
\writetoca{Appendix {#1.} {#2}}\par\nobreak\medskip\nobreak}
%
%
\def\eqnn#1{\xdef #1{(\secsym\the\meqno)}\writedef{#1\leftbracket#1}%
\global\advance\meqno by1\wrlabeL#1}
\def\eqna#1{\xdef #1##1{\hbox{$(\secsym\the\meqno##1)$}}
\writedef{#1\numbersign1\leftbracket#1{\numbersign1}}%
\global\advance\meqno by1\wrlabeL{#1$\{\}$}}
\def\eqn#1#2{\xdef #1{(\secsym\the\meqno)}\writedef{#1\leftbracket#1}%
\global\advance\meqno by1$$#2\eqno#1\eqlabeL#1$$}
%
\newskip\footskip\footskip14pt plus 1pt minus 1pt 
\def\footnotefont{\ninepoint}\def\f@t#1{\footnotefont #1\@foot}
\def\f@@t{\baselineskip\footskip\bgroup\footnotefont\aftergroup\@foot\let\next}
\setbox\strutbox=\hbox{\vrule height9.5pt depth4.5pt width0pt}
\global\newcount\ftno \global\ftno=0
\def\foot{\global\advance\ftno by1\footnote{$^{\the\ftno}$}}
%
\newwrite\ftfile
\def\footend{\def\foot{\global\advance\ftno by1\chardef\wfile=\ftfile
$^{\the\ftno}$\ifnum\ftno=1\immediate\openout\ftfile=foots.tmp\fi%
\immediate\write\ftfile{\noexpand\smallskip%
\noexpand\item{f\the\ftno:\ }\pctsign}\findarg}%
\def\footatend{\vfill\eject\immediate\closeout\ftfile{\parindent=20pt
\centerline{\bf Footnotes}\nobreak\bigskip\input foots.tmp }}}
\def\footatend{}
%
%
\global\newcount\refno \global\refno=1
\newwrite\rfile
\def\ref{[\the\refno]\nref}
\def\nref#1{\xdef#1{[\the\refno]}\writedef{#1\leftbracket#1}%
\ifnum\refno=1\immediate\openout\rfile=refs.tmp\fi
\global\advance\refno by1\chardef\wfile=\rfile\immediate
\write\rfile{\noexpand\item{#1\
}\reflabeL{#1\hskip.31in}\pctsign}\findarg}
\def\findarg#1#{\begingroup\obeylines\newlinechar=`\^^M\pass@rg}
{\obeylines\gdef\pass@rg#1{\writ@line\relax #1^^M\hbox{}^^M}%
\gdef\writ@line#1^^M{\expandafter\toks0\expandafter{\striprel@x #1}%
\edef\next{\the\toks0}\ifx\next\em@rk\let\next=\endgroup\else\ifx\next\empty%
\else\immediate\write\wfile{\the\toks0}\fi\let\next=\writ@line\fi\next\relax}}
\def\striprel@x#1{} \def\em@rk{\hbox{}}
\def\lref{\begingroup\obeylines\lr@f}
\def\lr@f#1#2{\gdef#1{\ref#1{#2}}\endgroup\unskip}
\def\semi{;\hfil\break}
\def\addref#1{\immediate\write\rfile{\noexpand\item{}#1}} 
\def\startrefs#1{\immediate\openout\rfile=refs.tmp\refno=#1}
\def\xref{\expandafter\xr@f}\def\xr@f[#1]{#1}
\def\refs#1{\count255=1[\r@fs #1{\hbox{}}]}
\def\r@fs#1{\ifx\und@fined#1\message{reflabel \string#1 is undefined.}%
\nref#1{need to supply reference \string#1.}\fi%
\vphantom{\hphantom{#1}}\edef\next{#1}\ifx\next\em@rk\def\next{}%
\else\ifx\next#1\ifodd\count255\relax\xref#1\count255=0\fi%
\else#1\count255=1\fi\let\next=\r@fs\fi\next}
\newwrite\ffile\global\newcount\figno \global\figno=1
\def\fig{fig.~\the\figno\nfig}
\def\nfig#1{\xdef#1{fig.~\the\figno}%
\writedef{#1\leftbracket fig.\noexpand~\the\figno}%
\ifnum\figno=1\immediate\openout\ffile=figs.tmp\fi\chardef\wfile=\ffile%
\immediate\write\ffile{\noexpand\medskip\noexpand\item{Fig.\
\the\figno. }
\reflabeL{#1\hskip.55in}\pctsign}\global\advance\figno
by1\findarg}
\def\xfig{\expandafter\xf@g}\def\xf@g fig.\penalty\@M\ {}
\def\figs#1{figs.~\f@gs #1{\hbox{}}}
\def\f@gs#1{\edef\next{#1}\ifx\next\em@rk\def\next{}\else
\ifx\next#1\xfig #1\else#1\fi\let\next=\f@gs\fi\next}
\newwrite\lfile
{\escapechar-1\xdef\pctsign{\string\%}\xdef\leftbracket{\string\{}
\xdef\rightbracket{\string\}}\xdef\numbersign{\string\#}}

\def\writestop{\def\writestoppt{\immediate\write\lfile{\string\pageno%
\the\pageno\string\startrefs\leftbracket\the\refno\rightbracket%
\string\def\string\secsym\leftbracket\secsym\rightbracket%
\string\secno\the\secno\string\meqno\the\meqno}\immediate\closeout\lfile}}
\def\writestoppt{}\def\writedef#1{}
\def\seclab#1{\xdef #1{\the\secno}\writedef{#1\leftbracket#1}\wrlabeL{#1=#1}}
\def\subseclab#1{\xdef #1{\secsym\the\subsecno}%
\writedef{#1\leftbracket#1}\wrlabeL{#1=#1}}
\newwrite\tfile \def\writetoca#1{}
\def\leaderfill{\leaders\hbox to 1em{\hss.\hss}\hfill}
\def\writetoc{\immediate\openout\tfile=toc.tmp
    \def\writetoca##1{{\edef\next{\write\tfile{\noindent ##1
    \string\leaderfill {\noexpand\number\pageno} \par}}\next}}}

%
\catcode`\@=12 
%
\edef\tfontsize{\ifx\answ\bigans scaled\magstep3\else
scaled\magstep4\fi} \font\titlerm=cmr10 \tfontsize
\font\titlerms=cmr7 \tfontsize \font\titlermss=cmr5 \tfontsize
\font\titlei=cmmi10 \tfontsize \font\titleis=cmmi7 \tfontsize
\font\titleiss=cmmi5 \tfontsize \font\titlesy=cmsy10 \tfontsize
\font\titlesys=cmsy7 \tfontsize \font\titlesyss=cmsy5 \tfontsize
\font\titleit=cmti10 \tfontsize \skewchar\titlei='177
\skewchar\titleis='177 \skewchar\titleiss='177
\skewchar\titlesy='60 \skewchar\titlesys='60
\skewchar\titlesyss='60
\def\titlefont{\def\rm{\fam0\titlerm}
\textfont0=\titlerm \scriptfont0=\titlerms
\scriptscriptfont0=\titlermss \textfont1=\titlei
\scriptfont1=\titleis \scriptscriptfont1=\titleiss
\textfont2=\titlesy \scriptfont2=\titlesys
\scriptscriptfont2=\titlesyss \textfont\itfam=\titleit
\def\it{\fam\itfam\titleit}\rm}
 \ifx\answ\bigans\else scaled\magstep1\fi
\ifx\answ\bigans\def\abstractfont{\tenpoint}\else
\font\abssl=cmsl10 scaled \magstep1 \font\absrm=cmr10
scaled\magstep1 \font\absrms=cmr7 scaled\magstep1
\font\absrmss=cmr5 scaled\magstep1 \font\absi=cmmi10
scaled\magstep1 \font\absis=cmmi7 scaled\magstep1
\font\absiss=cmmi5 scaled\magstep1 \font\abssy=cmsy10
scaled\magstep1 \font\abssys=cmsy7 scaled\magstep1
\font\abssyss=cmsy5 scaled\magstep1 \font\absbf=cmbx10
scaled\magstep1 \skewchar\absi='177 \skewchar\absis='177
\skewchar\absiss='177 \skewchar\abssy='60 \skewchar\abssys='60
\skewchar\abssyss='60
\def\abstractfont{\def\rm{\fam0\absrm}
\textfont0=\absrm \scriptfont0=\absrms \scriptscriptfont0=\absrmss
\textfont1=\absi \scriptfont1=\absis \scriptscriptfont1=\absiss
\textfont2=\abssy \scriptfont2=\abssys \scriptscriptfont2=\abssyss
\textfont\itfam=\bigit \def\it{\fam\itfam\bigit}\def\footnotefont{\tenpoint}%
\textfont\slfam=\abssl \def\sl{\fam\slfam\abssl}%
\textfont\bffam=\absbf \def\bf{\fam\bffam\absbf}\rm}\fi
\def\tenpoint{\def\rm{\fam0\tenrm}
\textfont0=\tenrm \scriptfont0=\sevenrm \scriptscriptfont0=\fiverm
\textfont1=\teni  \scriptfont1=\seveni  \scriptscriptfont1=\fivei
\textfont2=\tensy \scriptfont2=\sevensy \scriptscriptfont2=\fivesy
\textfont\itfam=\tenit \def\it{\fam\itfam\tenit}\def\footnotefont{\ninepoint}%
\textfont\bffam=\tenbf
\def\bf{\fam\bffam\tenbf}\def\sl{\fam\slfam\tensl}\rm}
\font\ninerm=cmr9 \font\sixrm=cmr6 \font\ninei=cmmi9
\font\sixi=cmmi6 \font\ninesy=cmsy9 \font\sixsy=cmsy6
\font\ninebf=cmbx9 \font\nineit=cmti9 \font\ninesl=cmsl9
\skewchar\ninei='177 \skewchar\sixi='177 \skewchar\ninesy='60
\skewchar\sixsy='60
\def\ninepoint{\def\rm{\fam0\ninerm}
\textfont0=\ninerm \scriptfont0=\sixrm \scriptscriptfont0=\fiverm
\textfont1=\ninei \scriptfont1=\sixi \scriptscriptfont1=\fivei
\textfont2=\ninesy \scriptfont2=\sixsy \scriptscriptfont2=\fivesy
\textfont\itfam=\ninei \def\it{\fam\itfam\nineit}\def\sl{\fam\slfam\ninesl}%
\textfont\bffam=\ninebf \def\bf{\fam\bffam\ninebf}\rm}
%
%

\hyphenation{anom-aly anom-alies coun-ter-term coun-ter-terms}
\def\inv{^{\raise.15ex\hbox{${\scriptscriptstyle -}$}\kern-.05em 1}}

\def\Dsl{\,\raise.15ex\hbox{/}\mkern-13.5mu D} 
\def\dsl{\raise.15ex\hbox{/}\kern-.57em\partial}

\font\bigit=cmti10 scaled \magstep1
\def\lspace{\ifx\answ\bigans{}\else\qquad\fi}
\def\lbspace{\ifx\answ\bigans{}\else\hskip-.2in\fi} 
\def\boxeqn#1{\vcenter{\vbox{\hrule\hbox{\vrule\kern3pt\vbox{\kern3pt
     \hbox{${\displaystyle #1}$}\kern3pt}\kern3pt\vrule}\hrule}}}
\def\mbox#1#2{\vcenter{\hrule \hbox{\vrule height#2in
         \kern#1in \vrule} \hrule}}  
%
 \def\CO{{\cal O}} 
  \def\CF{{\cal F}} 
\def\CL{{\cal L}}   
 \def\CR{{\cal R}}  
\def\e#1{{\rm e}^{^{\textstyle#1}}}

\def\darr#1{\raise1.5ex\hbox{$\leftrightarrow$}\mkern-16.5mu #1}

\def\half{{\textstyle{1\over2}}} 
\def\roughly#1{\raise.3ex\hbox{$#1$\kern-.75em\lower1ex\hbox{$\sim$}}}

\def\ap#1#2#3{Ann. Phys. {\bf #1} (#2) #3}

\def\jhep#1#2#3{JHEP {\bf#1}(#2) #3}

\def\ap#1#2#3{Ann.~Phys. {\bf #1} (#2) #3}
\def\IB{\relax\hbox{$\inbar\kern-.3em{\rm B}$}}
\def\IC{\relax\hbox{$\inbar\kern-.3em{\rm C}$}}
\def\ID{\relax\hbox{$\inbar\kern-.3em{\rm D}$}}
\def\IE{\relax\hbox{$\inbar\kern-.3em{\rm E}$}}
\def\IF{\relax\hbox{$\inbar\kern-.3em{\rm F}$}}
\def\IG{\relax\hbox{$\inbar\kern-.3em{\rm G}$}}
\def\IGa{\relax\hbox{${\rm I}\kern-.18em\Gamma$}}
\def\IH{\relax{\rm I\kern-.18em H}}
\def\IK{\relax{\rm I\kern-.18em K}}
\def\IL{\relax{\rm I\kern-.18em L}}
\def\IP{\relax{\rm I\kern-.18em P}}
\def\IR{\relax{\rm I\kern-.18em R}}
\def\IZ{\relax\ifmmode\mathchoice{
\hbox{\cmss Z\kern-.4em Z}}{\hbox{\cmss Z\kern-.4em Z}}
{\lower.9pt\hbox{\cmsss Z\kern-.4em Z}} {\lower1.2pt\hbox{\cmsss
Z\kern-.4em Z}} \else{\cmss Z\kern-.4em Z}\fi}
\def\II{\relax{\rm I\kern-.18em I}}


\def\CF{{\cal F}}

\def\CL{{\cal L}}
\def\CM{{\cal M}}
\def\CN{{\cal N}}
\def\CO{{\cal O}}

\def\CR{{\cal R}}
\def\CS{{\cal S}}

\def\CW{{\cal W}}

\def\p{\partial}

\def\dd{{\rm d}}

\def\zb{\bar{z}}
\def\la{\langle\ldots}
\def\ra{\ldots\rangle}


\def\inbar{\,\vrule height1.5ex width.4pt depth0pt}
\font\cmss=cmss10 \font\cmsss=cmss10 at 7pt

\input epsf.tex

\def\a{{\alpha}}
\def\ap{{\a}^{\prime}}

\def\d{{\delta}}
\def\g{{\gamma}}
\def\e{{\epsilon}}

\def\ve{{\varepsilon}}
\def\vf{{\varphi}}
\def\m{{\mu}}
\def\n{{\nu}}

\def\l{{\lambda}}
\def\s{{\sigma}}
\def\t{{\theta}}
\def\o{{\omega}}
\def\Om{{\Omega}}
\def\nc{noncommutative\ }

\def\G{{\Gamma}}
\def\Gt{{\tilde \Gamma}}
\def\P{{\Pi}}

\def\ss{{\bf s}}
\def\na{{\nabla}}
\def\nt{{\tilde\nabla}}
\def\MM{{\CM_{\Sigma,X}}}
\def\lref{\begingroup\obeylines\lr@f}
\def\lr@f#1#2{\gdef#1{\ref#1{#2}}\endgroup\unskip}

\lref\poisigm{P.~Schaller, T.~Strobl, hep-th/9405110, Mod. Phys. Lett. {\bf A9} (1994) 3129-3136\semi
N.~Ikeda, hep-th/9312059, Ann. Phys. {\bf 235} (1994) 435-464}

\lref\wittop{E.~Witten, ``Mirror symmetry and topological field
theory '',  hep-th/9112056 }

\lref\as{A.~Schwarz, "Geometry of Batalin-Vilkovisky quantization"
  Commun.Math.Phys. 155 (1993) 249-260}
\lref\aksz{ M. Alexandrov, M. Kontsevich, A. Schwarz, O. Zaboronsky,
"The Geometry of the Master Equation and Topological Quantum Field Theory",
      Int.J.Mod.Phys. A12 (1997) 1405-1430\semi
        A.S.~Cattaneo, G.~Felder
     "On the AKSZ formulation of the Poisson sigma model", math.QA/0102108}

\lref\maxim{M.~Kontsevich, ``Deformation quantization of Poisson
manifolds, I'', q-alg/9709040} \lref\verroc{M.~Rocek,
E.~Verlinde,"Duality, Quotients, and Currents"
      Nucl.Phys. B373 (1992) 630-646}
\lref\cf{A.S. ~Cattaneo, G.~Felder,
      "A path integral approach to the Kontsevich quantization formula"
     Commun.Math.Phys. 212 (2000) 591-611}
\lref\toappear{L.~Baulieu, A.~Losev, N.~Nekrasov, to appear}
\lref\manchon{D.~Manchon, "Poisson bracket, deformed bracket and
 gauge group actions in Kontsevich deformation quantization",
  Letters in Mathematical Physics {\bf 52}: 301-310 }
\lref\stress{A.~Gerhold, J.~Grimstrup, H.~Grosse, L.~Popp,
M.~Schweda, and R.~Wulkenhaar, hep-th/0012112 \semi Y.~Okawa,
H.~Ooguri, hep-th/0103124\semi M.~Abou-Zeid, H.~Dorn,
hep-th/0104244} \lref\old{L.~Baulieu, A.~Losev, N.~Nekrasov,
unpublished; talk by A.L. in Luminy, 1999} \lref\cfnew{
A.S.~Cattaneo, G.~Felder, L.~Tomassini, "From local to global
deformation quantization of Poisson manifolds",
     math.QA/0012228}

\lref\witsei{N.~Seiberg, E.~Witten, hep-th/9908142,
\jhep{9909}{1999}{032}}

\Title{\vbox{\baselineskip 10pt \hbox{ITEP-TH-74/00}
\hbox{IHES-NN02/01} \hbox{hep-th/0106042} }} {\vbox{\vskip -30
true pt
   \centerline{ TARGET SPACE SYMMETRIES}
 \smallskip\smallskip
   \centerline{IN  TOPOLOGICAL THEORIES I.}
\medskip \vskip4pt }} \vskip -20 true pt \centerline{Laurent Baulieu$^{1}$,
Andrei S.~Losev$^{1,2}$, Nikita A.~Nekrasov$^{2,3}$}
\smallskip\smallskip
\centerline{$^{1}$\it LPTHE, Universit\'e Paris VI, 4 Place
Jussieu, T.16,  $1^{er}$ etage, Paris 05 CEDEX France}
\centerline{$^{2}$\it Institute for Theoretical and Experimental
Physics, 117259 Moscow, Russia} \centerline{$^{2,3}$\it Institute
des Hautes Etudes Scientifiques, Le Bois-Marie, Bures-sur-Yvette,
F-91440 France}
\medskip \centerline{\rm e-mail: baulieu@lpthe.jussieu.fr,
lossev@gate.itep.ru,  nikita@ihes.fr}

\bigskip
We study realization of the target space diffeomorphisms in the
type $C$ topological string.
 We found
that the charges, which generate  transformations of the boundary
observables,  form an algebra, which differs from that of bulk
charges by the contribution of the bubbled disks. We discuss
applications to noncommutative field theories.

\newsec{Introduction}

One of the great achievements of string theory is the construction
of a quantum theory containing gravity. As such, the gauge
symmetry of general relativity -- space-time diffeomorphisms, or
their ${\ap}$-deformation must be present among the symmetries of
string theory. Of course, the study of this symmetry or its
deformation is obscured by a choice of a background metric
$\langle g_{\m\n} \rangle \neq 0$ which leaves only a
finite-dimensional group of isometries as explicit symmetry of the
problem.

However, as we shall see below, there are string theories, which
are formally related to a Seiberg-Witten ${\ap} \to 0$ limit (even
in the case of ordinary bosonic string) of a physical string,
which do not require target space metric at all. Instead, one
deals with Poisson tensors, and sometimes with connections (but
the dependence on the connection is in some sense trivial). The
choice of a background Poisson tensor $\langle {\t}^{\m\n} \rangle
\neq 0$ is much less restrictive as far as the group of
diffeomorphisms is concerned, for any Hamiltonian vector field
$V^{\m} = {\t}^{\m\n} {\p}_{\n} H$, where $H$ is a function on the
target space $X$, generates a symmetry of ${\t}$.

We are going to study these theories (they are called topological
strings of type $C$) and will show that the closed strings enjoy
the classical symmetry of Poisson diffeomorphisms, while the open
strings exhibit a non-trivial deformation of this symmetry. The
study of this deformed symmetry maybe a hint into what could be
happening with the physical string symmetries in generic
backgrounds.

The paper is organized as follows. In section $2$, we describe the
${\ap} \to 0$ limit and the unconventional branch of (topological)
string theories which emanate from this point. In section $3$, we
discover that in order to define the theory properly one is bound
to utilize the techniques of BV quantization. We find a solution
to BV master equation which enjoys target space covariance at the
expense of introducing a connection on the tangent bundle to $X$.
This new element appears upon careful examining of the properties
of the auxilliary fields needed to ensure the proper gauge fixing.
In section $4$, we continue our study of the target space
diffeomorphisms realized in the theory. We show that the closed
string symmetries  in general differ from those of open string.
The origin of this anomaly is traced back to the phenomenon of
``disk bubbling'', which is absent in the analogous quantum
mechanical models. In section $5$, we conclude by giving the
possible applications of the discovered symmetry to the
``covariant'' noncommutative field theories.

{\bf Remarks on notations.} Throughout the paper, we freely use
the notions of topological field theories, like $p$-observables,
ghost number, Witten's descend, $Q$-closedness etc, which are
introduced in \wittop. The target space coordinates are denoted by
$X^{\m}$ in section $2$, and $q^i$ in the rest of the paper.
Poisson bi-vectors are denoted by ${\t}^{\m\n}$ or ${\pi}^{ij}$.

{\bf Plans for the future.} We plan to write an extended paper
\toappear, which will contain the unified BV treatment of
topological strings of types $A, B, C$, operator approach to
topological (Hodge) quantum mechanics, more thorough treatment of
the target space symmetries of topological field theories (beyond
the dimensions $\leq 2$), and applications to the recent
formulation of superstring by Berkovits.

{\bf Acknowledgements.} We would like to thank M.~Douglas, D.~Gross, G.~Horowitz, M.~Kontsevich, and S.~Shatashvili for discussions. Research of A.L. and N.N. was partially supported by RFFI under the grant 00-02-16530
and by the grant 00-15-96557 for the support of scientific schools, that of A.L. in addition by INTAS under the grant 99-590.
A.L. would like to thank LPTHE at University Pierre et Marie Curie, Paris, IHES, Bures-sur-Yvette, and TPI at University of Minnesota for their hospitality
while preparing this manuscript. N.N. thanks ITP, UC Santa Barbara for hospitality and Clay Mathematical Institute for support.

\newsec{Seiberg-Witten limit and Poisson sigma model}

\subsec{Approaching from the physical side} Consider the action of
bosonic string in the generic background of massless fields
(without ghosts, see below): \eqn\bs{S = {1\over{4\pi\ap}}
\int_{\Sigma} {\rm d}^2 {\s} \ h^{\half} \left[  \left( g_{\m\n}
(X) h^{ab} + i {\e}^{ab} B_{\m\n}(X) \right)  {\p}_{a} X^{\m}
{\p}_{b} X^{\n} + {\ap} R^{(2)} (h) {\Phi}(X) \right]} where
${\Sigma}$ is the worldsheet Riemann surface with the metric
$h_{ab}$, $h = {\rm det} h_{ab}$, and local coordinates ${\s}^a,
a=1,2$, $g_{\m\n}$ is the metric on the target space $X$, with
local coordinates $X^{\m}$, and $B_{\m\n}$ is the two-form on $X$
(more precisely, $B$ is defined up to gauge transformations $B \to
B + {\Lambda}$, where ${\Lambda}$ is a closed two-from whose
periods are in $8{\pi}^2 {\ap} {\IZ}$ so that globally on $X$, $B$
needs not to be well-defined, and in fact couples to the
word-sheet action via a Wess-Zumino term). For \bs\ to describe a
conformal sigma model, the metric $g$, the $B$-field and the
dilaton field ${\Phi}$ have to solve the beta-function equations:
\eqn\bfun{\eqalign{& R_{\m\n} (g) + 2 {\nabla}_{\m} {\nabla}_{\n}
{\Phi}  - {1\over 4} H_{\m\l\o} H_{\n}^{\l\o} = O( {\ap} ) \cr &
{\nabla}^{\o} \left( e^{- 2\Phi} H_{\o\m\n} \right) = O ({\ap} )
\cr & {{D - 26}\over{6}} + {\ap} {\nabla}_{\o} {\Phi}
{\nabla}^{\o} {\Phi} - {{\ap}\over{2}} {\nabla}^{2} {\Phi}  -
{{\ap}\over{24}} H_{\m\n\l}H^{\m\n\l} = O ({\ap}^2)\cr}} where
$R_{\m\n}(g)$ is the Ricci tensor of $g$, and $H = dB$. Let us
re-write \bs\ in the first order form\foot{Note that this is not
the conventional passage to the Hamiltonian framework, where one
would have gotten a single component of $p$, thereby breaking two
dimensional covariance}. To this end introduce a one-form $p_{\m}$
on $\Sigma$, with values in $T^* X$: $p_{\m} = p_{\m, a} {\rm
d}{\s}^a$ and write an equivalent (after eliminating $p$) to \bs\
action: \eqn\bsfo{S = \int_{\Sigma} i \ p_{\m} \wedge {\rm d}X^{\m} +
{\pi}{\ap} G^{\m\n} (X) p_{\m} \wedge \star p_{\n} +
{\half}{\t}^{\m\n} (X) p_{\m} \wedge p_{\n} + \ dilatonic \ terms
} where $\star$ is the two dimensional Hodge star operation on
one-forms, which depends on $h^{\half} h^{ab}$, ${\star}^2 = -1$,
and \eqn\opst{\left( g +  B \right)^{-1} = G +
{{\t}\over{2\pi\ap}}} Imagine now taking the ${\ap} \to 0$ limit,
while holding $G$ and ${\t}$ fixed (Seiberg-Witten limit \witsei).
It means that \eqn\scal{\eqalign{& g \sim {(2\pi\ap)}^2 {1\over\t} G
{1\over{{\t}^{t}}} \cr & B \sim 2{\pi}{\ap} {1\over{\t}} \cr}}
From \bfun\ we may now derive the ${\ap} \to 0$ limit of the
beta-function conditions. We shall set ${\Phi} =0$ for simplicity.
Since $R_{\m\n}$ is invariant under the global rescaling of the
metric the first term in the Einstein equation is $O(1)$, while
the $H^2$ term scales as: ${\ap}^{-2}$, which forces $H = 0$. The
next two equations are then automatically obeyed, as ${\ap}H^2
\sim {\ap}^{-1}$ dominates over $(D-26)/6  \sim {\ap}^0$ (at this
point we assumed that ${\t}$ is invertible).

Thus, we are approaching from the physical string side 
the ``theory'' with the action \eqn\acnaiv{S =
\int_{\Sigma}  i \ p_{\m} \wedge dX^{\m} + {\half} {\t}^{\m\n} p_{\m}
\wedge p_{\n}} where ${\t}$ is such that $d {\t}^{-1} = 0$. The
last equation (for invertible ${\t}$) implies that ${\t}$ is a
Poisson tensor, i.e. if one defines a bracket on the functions on
$X$ by the formula \eqn\posbr{\{ f, g \} = {\t}^{\m\n} (X) {{\p
f}\over{\p X^{\m}}} {{\p g}\over{\p X^{\n}}}}then it obeys Jacobi
identity: \eqn\jacb{\{ \{f, g \}, h\} + \{ \{ g, h \}, f\} + \{ \{
h, f \}, g \} = 0 .}

\subsec{Approaching from the topological side}

Now imagine that we started (as in \poisigm) with \acnaiv\ where ${\t}$ is not
necessarily an invertible Poisson tensor, i.e. ${\t}^{\m\n}
{\p}_{\n} {\t}^{\l\o} + \ cyclic \ permutations = 0$. The theory
with the action \acnaiv\ has a symmetry descending from that of
\bs\ - that of diffeomorphisms of $X$. It acts on $p_{\m}$ as on
the one-form on $X$, i.e. for the infinitesimal diffeomorphism,
\eqn\infd{{\d}X^{\m} = v^{\m}(X), \ {\d} p_{\m} = -  p_{\n}
{\p}_{\m} v^{\n}} Of course, the presence of $g$ and $B$ in \bs\
made them transform, thus making only a finite-dimensional
subgroup of $Diff(X)$ a symmetry, the rest acting on the space of
backgrounds. Similarly, the presence of ${\t}$ in \acnaiv\ reduces
$Diff(X)$, but this time to an infinite-dimensional group $PDiff
(X, {\t})$ of Poisson diffeomorphisms. Any \infd\ with $v^{\m} =
{\t}^{\m\n}{\p}_{\n} H$ for $H$  a function on $X$ generates a
symmetry of \acnaiv.

In addition, the action \acnaiv\ has a gauge symmetry \cf:
\def\dg{{\d}_{\ve}}
\eqn\gauges{\eqalign{& {\dg} p_{\m} = d {\ve}_{\m} -
{\p}_{\m}{\t}^{\l\o} {\ve}_{\l} p_{\o} \cr & {\dg} X^{\m} =
{\t}^{\m\n} {\ve}_{\n} \cr }} One can check that global symmetries
are incompatible with the local symmetries. One can fix that by
adding to ${\dg} p_{\m}$ the terms like ${\Gamma}_{\m\n}^{\l}
{\ve}_{\l} ( d X^{\n}  - {\t}^{\n\o} p_{\o} )$, which depend on
connection in tangent bundle to $X$ and make everything covariant
but then these transformations don't form a closed algebra
($Q_{BRST}$ is not nilpotent). It is the Batalin-Vilkovisky (BV)
formalism that saves the day, as we will see below.

{\bf Note.} We do not claim that the bosonic string and the type
$C$ string are equivalent or continuously connected. They are
clearly different ways of treating the ill-defined theory \acnaiv.

In the topological string we shall concentrate upon, the symmetry
$\dg$ is considered as a gauge symmetry and must be fixed. Also,
to get back the amplitudes of the physical string one must fix the
two-dimensional reparametrization invariance, which would add
$b-c$ ghost system, fix  Weyl invariance, while the gauge
invariance \infd\ is broken explicitly by the coupling to the
target space metric $G$. In the topological string the $b-c$
system is not needed -- BV machinery will contain all the
necessary ghosts and one can couple the system to the two
dimensional topological gravity. At genus zero, which is what we
shall study in this paper, this amounts to considering the
integrals of the vertex operators over a compactification of the
moduli space of points on the sphere (disk) up to the action of
the group $SL_2({\IC})$ ( $SL_2({\IR})$). The Feynman rules of
\maxim\cf\ automatically produce closed differential forms on
these spaces.

\newsec{Type $C$ topological sigma model}

\subsec{A snapshot of BV formalism} Here we view the BV formalism (see, cf. \as)
as an integral $I_{BV}$ of the BV differential form $\Om_{BV}$
along the Lagrangian submanifold $\CL$ in the BV space:
\eqn\maineq{I_{BV}=\int_{\CL} \Om_{BV}} The BV space $\CM$  is
equipped with the canonical odd symplectic form ${\o}_{BV}$. One
can choose local coordinates to identify ${\CM}$ with ${\Pi}T^* N$
where $N$ is some (super)manifold, where the symplectic form has a
canonical form \eqn\bvform{ \o_{BV}=\d Z_{a}^{+} \wedge \d Z^a }
where $Z^a$ denotes the (super)coordinates on $N$ and $Z_{a}^{+}$-
corresponding coordinates on the cotangent fiber.

The submanifold $\CL$ is Lagrangian with respect to the
canonical form $\o_{BV}$ (in the physical literature its generating function is called the gauge fermion).

The BV differential form  $\Om_{BV}$ is constructed out of two ingredients \as:
the BV action $S$ and the BV measure $\nu$: ${\Om}_{BV} = \left( {\n} e^{- S} \right)$.

The action $S$ must obey the so-called BV master equation:
\eqn\bvm{\{ S, S\}_{BV} := {\o}_{BV}^{-1} ({\p}_l S \wedge {\p}_r
S) = 0}

One calls the coordinates $Z^a$ the fields and $Z^{+}_{a}$ the anti-fields.
Sometimes one distinguishes the classical part of $N$ and the auxilliary fields used for gauge fixing. Also, the identification of the BV phase space
with ${\Pi} T^* N$ is not unique and is not global in general, so the partition of all the fields involved on the fields and anti-fields is not unique.

The deformations of the action $S$ that preserve \bvm\ are (in the
first order approximation) the functions ${\Phi}$ on $\CM$ which
are $Q_{BV}$-closed, where the differential $Q_{BV}$ acts as
$Q_{BV} {\Phi} = \{ S , {\Phi} \}_{BV}$. The deformations which
are $Q_{BV}$-exact are trivial in the sense that they could be
removed by a symplectomorphism of ${\CM}$ (one has to make sure
that this symplectomorphism preserves $\n$ to guarantee that the
quantum theory is not sensitive to such a $Q_{BV}$--exact term).

\subsec{Back to the $\int p dX$ theory}

In this section we shall embed the action \acnaiv\ into the BV framework.
We start with the ${\t}= 0$ case. We shall change the notations compared with the physical string case - the coordinates on $X$ will be denoted mostly as
$q^i$.

Consider the space $\MM$ of maps of supermanifolds \aksz:
$$
{\MM} = {\rm Maps} ( \P T\Sigma, \P T^{*}X )
$$
If we choose on $T^*X$ the coordinates $(p_i, q^i)$, $i=1, \ldots,
{\rm dim}X$, with $q^i$ being the coordinates
on $X$, then ${\vf} \in \MM$ can be expressed via
the following objects:
\def\da{\dagger}
\eqn\oups{ {\vf}^* q^i = 
Q^{i} = q^{i}_{(0)} + q^{i}_{(1)} + q^{i}_{(2)} , \quad {\vf}^8 p_i = Q^\da _{i} =
p_{i(0)} + p_{i(1)} + p_{i (2)}
}
where the component with a subscript $(a)$ is a $a$-form on $\Sigma$, valued insome fiber bundle over $\Sigma$ (the map from $\Pi T\Sigma$ is a collection of differential forms on $\Sigma$). In
this expansion the pairs of field and antifields are just given by the
pairs  $(Q^{i}_{(a)}  , Q^{\da}_{i (2-a)} )$ for each value
$0\leq a\leq 2$.

In quantum field theory, $q^i=q^{i}_{(0)}$ and $p_i=p_{i(1)}$ are
the classical fields present in the classical
 action, which is obtained from the BV action by setting all non-classical
fields to zero. The component $q^i_{0}$ describes the ordinary map from $\Sigma$ to $X$.

Now consider the effect of 
the change of coordinates 
\eqn\cco{
q^i \mapsto {\tilde q}^j (q), \ p_i \mapsto {\tilde p}_i = p_j {{\p q^j}\over {\p {\tilde q}^i}}
}
on $p_{i(a)}$ and
$q^{i}_{(a)}$. Here ${\p q}\over {\p {\tilde q}}$ is understood as the Jacobian of the inverse change of coordinates.

From the definition of our expansions we readily compute:
\eqn\transfi{\eqalign{& {\tilde Q}^j = {\tilde q}^j(Q) =
{\tilde q}^j(q_{(0)})
 + {{\p {\tilde q}^j}\over{\p q^i}} q^{i}_{(1)}
+ {{\p {\tilde q}^j}\over{\p q^i}} q^{i}_{(2)} +
{1\over 2} {{\p^2 {\tilde q}^j}\over{\p q^k \p q^l}}
q^{k}_{(1)}q^{l}_{(1)}\cr
}}
and
\eqn\transfip{\eqalign{
& {\tilde Q^\da}_{j} =
Q^\da_{i}
 {{\p  Q^i}\over{\p \tilde Q^j}}  \cr
}}
implies
\eqn\transfipp{\eqalign{&
  {\tilde p}_{j(0)} =
p_{i(0)}
{{\p q^i}\over{\p \tilde q^j}}  \cr
& {\tilde p}_{j(1)} =
p_{i(1)}  {{\p q^i}\over{\p \tilde q^j}}
 -
{{\p q^m}\over{\p \tilde q^l}}
{{\p q^n}\over{\p \tilde q^j}}
{{\p^2 \tilde q^k}\over{\p q^m \p q^n}}  p_{k (0)} q_{(1)}^l \cr}}
and we shall not need the explicit formula for ${\tilde p}_{j(2)}$
which is obtained by a straightforward tedious computation.

This transformations will be needed when adding to our systems of fields
the BRST quartets that are needed for achieving the gauge fixing of the
action \acnaiv. We will obtain  a set of  BRST
transformations corresponding to a BV system of rank two.  For such a
system, the non linear antifield dependence   forbids the use of the
familiar    Faddeev-Popov formula. It generally leads to   a ghost
and antighost dependence which is at least cubic. This makes the use of
antifields unavoidable and  will  {\it a fortiori}
justify   the use of the BV formalism.

\subsec{ The W-deformation of the AKSZ action}

Consider the action functional \aksz:
\eqn\bvi{S = \int_{\Sigma} Q^{\da}_i {\dd} Q^i + {\CW}(Q^{\da}, Q)}
where ${\CW} (Q^{\da}, Q)$ is a (target-space scalar)
function evaluated on the superfields $Q$ and $Q^{\da}$.
The integral in \bvi\ picks out  the
two-form component  $( Q^{\da}_i {\dd} Q^i +{\CW}(Q^{\da},
Q))|_{z\bar z}$. It is easy
to verify that ${\CW} (p_{(1)},q_{(0)})$ gives \acnaiv.  However the full content contains more
information, and the superfield formalism with ghost unification simplifies tremendously all
the formulae,
as well as their geometrical interpretation.

${S}$ must obey  the   BV master equation: \eqn\clas{\{ {S}, {S}
\} \equiv \sum_{i, a} {{{\d}^{r} S}\over{{\d p_i}^{(a)}}}
{{{\d}^{l} S}\over{{\d q^i}^{(2-a)}}} - (-)^{a} ( l
\leftrightarrow r) = 0}
\def\NN{{\CN}_{{\Sigma}, X}}
This implies that $\CW$ obeys:
\eqn\clasw{{{{\d} {\CW}}\over{{\d p_i}}}
{{{\d} {\CW}}\over{{\d q^i}}}
- (-)^{a} ( l \leftrightarrow r) = 0}

In the example that we  will study shortly in some detail
\eqn\defofpi{ {\CW} = {1\over 2} \pi^{ij} (Q) Q_{i}^{+}Q_{j}^{+} } and the condition
\clasw\ is equivalent to the statement that the bi-vector $\pi$ is
of Poisson type \jacb\ (we changed the notation ${\t}^{ij} \to {\pi}^{ij}$).

\subsec{Quartets and $Diff(X)$}

Let us now present the  improvements  needed in order
to make  the target space covariant  gauge-fixing of the theory.
 Eventually,     antighosts
and  Lagrange multipliers are needed for the gauge-fixing. We will
generically denote them as  $\chi$ and $H$, respectively. We thus
consider the space $\NN$ which is a fiber bundle  over $\MM$ with
the fiber spanned by  the quadruples $(\chi^i, H^i, \chi_i^{+},
H_{i}^{+})$, where $\chi^i, H^i$ are fermionic and bosonic
zero-forms on $\Sigma$ with values in $q^{*}TX$  (as usual $q^*$ denotes 
the pull-back with respect to $q_{(0)}$).

As an example \toappear, if $ \chi^i$ and $  H^i$ are fermionic
and bosonic zero-forms on $\Sigma$, then $\chi_i^{\da}$ and $
H_i^{\da}$ are respectively   bosonic and fermionic two-forms with
values in $q^{*}T^* X$.

Let us first show that the correct transformation
law for the superfields
$Q^{\da}, Q, \chi, $ and so forth
must be modified in the presence of anyone of the quartets $(\chi^i,
H^i, \chi_i^{+}, H_{i}^{+})$.
We will find that the reparametrization invariance must be modified into:
\eqn\ctr{\eqalign{& \tilde Q^{\da}_l
 = Q^{\da}_i {{\p Q^i}\over{\p \tilde Q^l}} - {{\p q^j}\over{\p \tilde q^i}}{{\p
q^n}\over{\p \tilde q^l}} {{\p^2 \tilde  q^i}\over{\p q^m \p q^n}} \left(
\chi^{+}_j \chi^m + H^{+}_j H^m \right) \cr \quad & \tilde\chi^i = \chi^j {{\p
\tilde q^i}\over{\p q^j}},\quad  \tilde\chi^{+}_i = \chi^{+}_{j}{{\p
q^{j}}\over{\p \tilde  q^{i}}} \cr & \tilde H^i =  H^j  {{\p \tilde q^i}\over{\p
q^j}},\quad  \tilde H^{+}_{i}=  H^{+}_{j}{{\p q^{j}}\over{\p \tilde q^{i}}} \cr}} In
the first formula, one can use $Q, \tilde Q$ instead of $q, \tilde
q$,  since
  $\chi^{+}_j
\chi^n + H^{+}_j H^n $ is always a
 two-form that  automatically projects
down to the $0$-th component of the superfield $Q$.

The necessity of defining  the coordinate  transformations  as in
\ctr\ is that we need a coordinate-invariant symplectic form on
the space $\CM$ for possibly defining   a covariant path integral
after the introduction of the fields $\chi$ and $H$.   Indeed,
$\NN$ is endowed with the odd symplectic form, which is invariant
under \ctr: \eqn\osf{\Omega = \int_{\Sigma} {\d} Q^{\da}_i \wedge
\d Q^i + \d
 \chi^i \wedge \d \chi_i^{+} + \d H^{+}_i \wedge \d H^{i}}

\subsec{Naive BV action}

As a first try,  we assume that the BV action is equal to:
\eqn\inc{
{\CS}^{\rm naive} =
\int_{\Sigma} Q^{\da}_i {\dd}Q^i + {\CW} (Q^{\da},Q) +
 \chi_i^{+} H^{i} }
${\CS}^{\rm naive} $ obeys the equation \clas\ if $S$ does.
It  induces the following
Hamiltonian vector field action
$\ss =
\{\CS ^{\rm naive}, {\bf .}\}$ on the space $\NN$:
\eqn\brst{\eqalign{\ss Q^{\da}_i = {\dd} Q^{\da}_i  - {{\p \CW}\over{\p
X^i}},
\quad &
\ss Q^i = {\dd} Q^i + {{\p \CW}\over{\p Q^{\da}_i}} \cr
\ss \chi^i = H^i, \quad & \ss \chi_i^{+} = H_{i}^{+} \cr
\ss H^i = 0, \quad & \ss H^{+}_{i} = 0\cr}}$\ss$ is nilpotent
due the fact that
we have introduced the ${\chi}, H$
dependence without spoiling the BV master equation.
The point is that it  is   necessary to define the transformation
property of
$Q^{\da}$ as in \ctr\ in order that  \osf\ be invariant. But then, the
action
${\CS}^{\rm naive}$  \inc,  which satisfies the BV
equation $\{{\CS}^{\rm naive},{\CS}^{\rm naive}\}=0$,
is not invariant under \ctr.

\subsec{Modified action}

To solve this contradiction, we must modify
${\CS}^{\rm naive}$ into a new action. By trial and error,
one finds an action
${\CS}_\g$ that must  explicitly depend  on the  choice of  a
connection  $\g^i_{jk}(q^{(0)})$  on the tangent bundle $TX$ to $X$.
This modified action
will be  covariant with respect to \ctr\ and still obey
$\{ {\CS}_\g, {\CS}_\g \}=0$.
The crucial
subtlety   is thus that  one needs additional terms in order  that
the function
${\CW}(Q^{\da},Q)$
  be  coordinate-independent.  This  is a non-trivial
requirement, given the intricate formula \ctr.

As a first attempt, one adds to $ {\CS}^{\rm naive}$ the term:
\eqn\add{
{{\p {\CW}}\over{\p Q^{\da}_i}}
\g^{j}_{ik} \left( \chi^{+}_j \chi^k +
H^{+}_j H^k \right) }
When one checks if this modified action
obeys the master equation, one finds that it requires corrections that
are non linear in $\g$.  So, one needs higher order corrections to the
action.    Fortunately the
procedure stops here with the following result:
\eqn\fllbv{\eqalign {{\CS}_{\g} = \int_\Sigma  Q^{\da}_i {\dd} Q^i +
\chi^{+}_i H^i + & {\CW} (Q^{\da}, Q)
+
{{\p {\CW}}\over{\p Q^{\da}_i}}
\g^{j}_{ik} \left( \chi^{+}_j \chi^k +
H^{+}_j H^k \right) \cr &
+ R({\g})^{i}_{jkl}  {{\p {\CW}}\over{\p Q^{\da}_j}} {{\p {\CW}}\over{\p
Q^{\da}_k}} {\chi}^l H^{+}_i }}
In order to prove that ${\CS}_{\g}$ obeys \clas, we just need to prove the following:
\eqn\nik{{{{\d} {\CS}_{{\g} + t {\a}}}\over{{\d}t}}_{t=0}
= \{ {\CS}_{\g}, {\CR}_{\a} \}}
where
\eqn\niki{{\CR}_{\a} = \int {{\p \CW}\over{\p Q^{\da}_i}} {\a}^{j}_{ik}
{\chi}^i H^{+}_j }
As a consequence, we have that for any value of the connection $\g$,  $\{ {\CS}_{\g}, {\CS}_{\g}
\} = 0$.
The proof of \nik\ is a simple computation. In particular,
for $\g = 0$ the statement is trivial given
\clasw. For $ t\sim 0$ it is also simple since
the last term in \fllbv\ can be neglected.

Given \nik\ the Poisson bracket
$\{ {\CS}_{\g}, {\CS}_{\g} \}$ is a solution to the
first order differential equation in $\g$ and hence vanishes in the
light of the initial condition $\{ {\CS}_{0}, {\CS}_{0} \} = 0$.

\subsec{Boundary conditions}

Here we
specify the boundary conditions on the fields
$(Q^i, Q_i^{\dagger} \chi^i, H^i, \chi_i^{+}, H_i^{+})$, following \cf:
\eqn\bndcnd{
\eqalign{\star Q^i \vert_{\P T\p \Sigma} = 0, \quad &
\quad Q_i^{\dagger} \vert_{\P T\p \Sigma} = 0 \cr
{\dd} \chi^i \vert_{\p \Sigma} = 0, \quad & \quad H^i \vert_{\p \Sigma} = 0\cr
{\dd} \star \chi_i^{+} \vert_{\p \Sigma} = 0, \quad & \quad
\star H_{i}^{+} \vert_{\p \Sigma} = 0 \cr}}

\subsec{$Diff(X)$ covariant gauge fixing of the C model}
We now turn to the construction of an explicit gauge,
that is, of a Lagrangian submanifold ${\CL} \subset {\NN}$.

We first choose a function $\Psi$ of half of the variables (the BV
gauge function of ghost number $-1$), which we can call symbolically $Z$,
 which is going to be
well-defined on ${\CL}$ and such that \eqn\ggfx{ {\d} {\Psi} =
{\d}^{-1} {\omega_{BV}} \vert_{\CL}, \qquad i.e. \ Z^{\da} = {{\d}{\Psi} \over {\d} Z}} 
We find it convenient to
perform a canonical transformation $(Q,Q^{\da}) \to (\rho, \xi)$:
\eqn\rota{\eqalign{ \xi^{i}_{(0)} = q^{i}_{(0)},\quad & \quad
\rho_{i(0)} = p_{i(0)} \cr \xi^{i}_{(1)} = q^{i}_{(1)}, \quad &
\quad \rho_{i(1)} = p_{i(1)} - {\G}_{ik}^{j} q^{k}_{(1)} p_{j(0)}
\cr \xi^{i}_{(2)} = q^{i}_{(2)} + {1\over 2} {\G}^{i}_{jk}
q^{j}_{(1)} q^{k}_{(1)}, \quad & \quad \rho_{i(2)} = p_{i(2)} -
{1\over 2}{\p}_i {\G}^{l}_{jk} p_{l(0)} q^{j}_{(1)} q^{k}_{(1)} \cr}} with
the virtue that ${\xi}_{(0,1)}, {\rho}_{(0,1)}$ transform homogeneously under the
coordinate transformations, unlike, say, $p_{i(1)}, p_{i(2)},
q_{(2)}^i$. ${\Gamma}$ has to be a torsion-free connection on $TX$
for \rota\ to be canonical.
Denote
$$
\na = {\dd}\xi^{l}_{(0)} \na_{l}, \quad
\na_{l} = \p_l + \G_l
$$
and similarly for $\nt$.
We choose the  following BV gauge
function $\Psi  ( \rho_{i(0)}, \rho_{i(1)}, \xi^{i}_{(0)},
\chi^i, H^i  )$ of the form:
\foot{Notice that we have chosen that     $\Psi$
does not depend on   $H$ or $\chi^{+}$.
Otherwise we would have to introduce a metric $G_{ij}$
on $X$. Our point is that we don't need to use metric,
the connections ${\Gamma}, {\Gamma}^{\prime}, \ldots$ suffice.
However, in \toappear\ we will elaborate on the effect of a linear $H$
dependence of $\Psi$ which
establishes a correspondence with A and B type models.}
\eqn\gfr{\Psi = \int_{\Sigma} \left( {\dd} \chi^{i} +
{\Gt}^{i}_{jk} \chi^{k} {\dd}
\xi^{j}_{(0)} \right) \star \rho_{i(1)} = \int_{\Sigma}
\chi^{i} {\nt} \star \rho_{i(1)}}
$\Psi$ being $H$ independent eliminates the Riemann tensor
dependence from the action, since the BV constraint \ggfx\
gives  that on   $\CL$:
\eqn\gfx{\eqalign{H^{+}_{i}  = 0, \quad & \quad
\chi_i^{+}  = - \nt \star  \rho_{i(1)} \cr
\xi^{i}_{(1)}  = \star \nt \chi^i, \quad & \quad
\xi^{i}_{(2)}  = 0  \cr
\rho_{i(2)}  = \p_i \Gt_{jk}^{l} \chi^k {\dd} \xi^{j}_{(0)}
& \star \rho_{l(1)}
- {\dd} \left( {\Gt}_{ij}^k \chi^j \star \rho_{k(1)} \right) \cr }}
To simplify the notations in what follows we re-define:
\eqn\chnno{\eqalign{
q^i = &  \xi^{i}_{(0)} \cr
p_i = & \rho_{i(1)} \cr
{\t}_i = & \rho_{i(0)} \cr}}
As before, we take:
\eqn\pois{{\CW}(Q, Q^{+}) = {1\over 2} \pi^{ij}(Q) Q^{+}_i Q^{+}_j}
with $\pi^{ij}$ being a Poisson bi-vector, i.e. bi-vector
obeying \jacb.
On ${\CL}$ the original fields-antifields are:
\eqn\faf{\eqalign{&Q^i =   q^i + \star \nt {\chi}^i - {\half} {\G}^{i}_{jk} \star \nt {\chi}^j \star \nt \chi^k \cr
& Q_i^{\da} = {\t}_i + p_i + {\G}^{j}_{ik} {\t}_j \star \nt \chi^k + \cr
& \qquad\qquad +  \half {\p}_i {\G}^l_{jk} {\t}_l \star \nt \chi^j \star \nt \chi^k + {\Gt}_{ik}^l \left( p_l \star \nt \chi^k + {\chi}^k \nt \star p_l \right) \cr 
 & \qquad \qquad + {\tilde R}^l_{ijk} {\chi}^k dq^j \star p_l \cr}}  
Let us make the final adjustment of the notations: introduce:
\eqn\hsmall{h^i = H^i + {\chi}^k \left( {\Gt} - {\g} \right)^i_{jk} {\pi}^{jl} {\t}_l}  
Note that the formula for $Q^i$ can be intepreted as an equation for 
a formal geodesic in $X$ with respect to the connection ${\G}$, 
which starts at the point $q$ along the formal tangent vector
$\star \nt \chi$. It is plausible that the similar relations will hold in the higher-dimensional analogues of the type $C$ sigma models.    
The restriction of the action functional $\CS$ on $\CL$ (= the gauge
fixed action) is given by (note that $\g$ has totally disappeared from the final Lagrangian):
\eqn\gf{\eqalign{& {\CS}_{\rm gf} = \int_{\Sigma}
p_i ( {\dd} q^i + \star \nt h^i) +  {\t}_i {\na} \star {\nt} \chi^i + \cr
& + {1\over 2} {\pi}^{ij} p_i p_j + \star \nt \chi^k \nt_k {\pi}^{ij} p_i {\t}_j  \cr
& + {\tilde R}_{ijk}^{l} {\pi}^{im} {\chi}^k {\t}_m dq^j \star p_l  - {\t}_i {\t}_j  \nt \chi^a  \nt \chi^b \left(  {1\over 2} {\na}^2_{ab} {\pi}^{ij}  - R_{cab}^{i} {\pi}^{cj} \right) \cr}   }
The beauty of our action  \fllbv\ seems lost when components are made explicit, and the built-in
 invariances seem awkward under this form, yet the coordinate covariance is manifest.
Also,  \gf\ shows
that we have a system of rank two,
due to the term
${\p}^2 {\pi} \nt{\chi}\nt{\chi} {\t}{\t}$ which is quadratic in
the antifields $q^{i}_{(1)}$.
Since the ${\Psi}$ is
$p_{(1)}$-dependent, this term  gives a non-trivial quartic ghost dependence of the gauge-fixed action. The later
  cannot be generated by some kind of Faddeev-Popov
determinant, which is the justification for the whole BV
machinery. Such terms are different in nature from those that one
obtains in the  $A$ model through the curvature dependent terms.
The $C$ model is thus quite different of the A-model, for which
the gauge function is independent on $p$, which implies that it
can be analyzed in the usual BRST formalism as a first rank
system.

\subsec{Covariant deformation quantization}

The final action \gf\ has the virtue of being a sigma model action, i.e. it is well-defined in terms of the geometrical data, i.e. the maps of the worldsheet into the target space endowed with the connections $\G, \Gt$ 
on its tangent bundle, as well as the Poisson bi-vector field $\pi$. 
It is not obvious that the action \gf\ defines a (super)conformal field theory for non-flat $\G, \Gt$, however, the RG flow, if any, should not affect the correlation functions of the 
$Q$-invariant observables. Thus one expects that the correlation function:
\eqn\crfnsp{
\langle f_1(q(0))  \ f_2 (q(1)) \rangle_{q(\infty) = q} =: f_1 \star f_2 (q)
}
defines an associative star-product. This star-product  will, nevertheless, depend on the connection
$\Gt$, even though all $\Gt$ dependence comes through $Q$-exact terms. 
Let us see how this can happen. We have no other way of treating the theory defined by \gf\ but by perturbation expansion. In order to generate the perurbation series we choose a classical solution $q (z, {\zb} ) = q, p = 0, \chi = 0, \t = 0, \ldots$, and expand around it. We should also keep track of the covariance properties of our expansion. As is standard in the sigma model techniques, 
\lref\tseytlin{A.A. Tseytlin, Phys.Lett.{\bf B} 178 (1986) 34}
e.g. \tseytlin, it is convenient to use the locally geodesic coordinates, which identify the vicinity of the point $q$ in the target space with the vicinity
of zero in the tangent space to the manifold at this point:
\eqn\geod{\eqalign{& q^i (z, {\zb}) = q^i +  {\bf y}^i - {\half}
{\Gt}^i_{jk}(q) {\bf y}^j {\bf y}^k + \ldots, \qquad {\bf y}^i = {\bf y}^i(z, \zb) \cr
& p_i (z, {\zb}) = {\bf p}_i + {\Gt}^{j}_{ik} (q){\bf y}^k {\bf p}_j + \ldots, \qquad\cr
& h^i (z, {\zb}) = {\bf h}^i - {\Gt}^i_{jk} (q) {\bf h}^j {\bf y}^k + \ldots\cr}}
and similarly for fermions. 
At this point we assume that $\Gt$ is also torsion-free. We can then set 
 $\G = \Gt$ for simplicity. Then the action
\gf\ becomes an infinite expansion in $\bf y$'s, with each vertex being constructed out of covariant expressions, like the curvatures, 
their covariant derivatives, 
the covariant derivatives of ${\pi}$ and so on, all at the point $q$. 
Similarly, 
the boundary observables $f (q(0))$ etc. become the expansions in 
$\bf y$ whose coefficients are 
nothing but the iterated (and symmetrized) covariand derivatives of $f$ at the point $q$.
The action now has the form:
$$
S = S_{0} + I  \ldots, 
$$
where 
$$
S_{0} = \int {\bf p}_i d{\bf y}^i + {\bf p}_i \star d {\bf h}^i + fermions
$$
and $I$ is the rest (containting $\pi$). We now expand $e^{-S}$ in $I$.
The important feature of the interaction density ${\CL}$, 
$ I = \int {\CL} $ is that it obeys the descend relations $
\{ Q, {\CL} \} = d{\CO}^{(1)}$ for some $1$-forms ${\CO}_k$.

Let us vary the correlation function
\crfnsp\ with respect to $\Gt$. It brings down the extra terms of the form $
\int \{ Q, {\d} R \} , R = {\nabla}_i{\d \Gt}^j_{kl}  {\bf y}^i d{\bf y}^k \star {\bf p}_j {\bf \chi}^l  + \ldots\}$. Normally we would use the fact that $Q$ is a symmetry of the theory to pull $Q$ off this term to make it act on the rest of the correlation function. This operation will convert the integrals
$$
{1\over k!} \int \langle \{ Q, R \} (z)  {\CL} (w_1) \ldots {\CL} (w_k) \rangle
$$
into
$$
{1\over k!} \int \langle R (z) \sum_{i =1}^{k} d {\CO} (w_i) 
\prod_{j \neq i} {\CL} (w_j) \rangle
$$
The total derivative $d{\CO}$ would make the integrated correlation
function vanish if  there were no boundaries in the integration domain.
There are two kinds of boundaries: $w_i \to w_j$ and $w_i \to $ the boundary of the worldsheet. We argue in \toappear\ that these boundary contributions are non-zero, thus providing the mechanism for non-decoupling of the $Q$-exact terms, needed for covariantization.

In the next subsections we shall not keep these $\G$-dependent terms, instead we shall analyze the currents and the charges generating the target space diffeomorphisms and will see the origin of the non-covariance of the original star product from a slightly different yet related perspective.

\newsec{The symmetries of bulk and boundary theories}

\subsec{Generalities on quantum symmetries in the bulk}

Recall that, in ordinary theories, the quantum global symmetry is
generated by a current $J_{V}$ that is conserved \eqn\ordin{ d
J_{V}=0}(inside correlation functions). Moreover the quantum
action of the global symmetry on the observable $\CO(x)$ inserted
at point $x$ is given by the  insertion of the expression $$
\int_{S(x,r)} J_V $$ under the correlator: \eqn\ordact{\lim_{r \to
0}  \la \int_{S(x,r)} J_{V} {\CO} (x) \ra = \la {\d}_{V} {\CO}(x)
\ra .} Here $S(x,r)$ is the set of points $y$ such that the
distance between $x$ and $y$ equals $r$, and $r \to 0$.

If the BV action and the measure are invariant under a global
symmetry, but the gauge fixing, that is a choice of Lagrangian
submanifold ${\CL}$ is not invariant, then the current is
conserved up to the $Q_{BV}$ exact terms, i.e. \eqn\bvcons{
dJ_{V}=  Q_{BV} ( J_{V,(2)} )  } This current is $Q_{BV}$--closed,
up to a total derivative, i.e.: \eqn\bvqclo{ Q_{BV}(J_V)= d J_{V,
(0)} } Note, that the above equation means that current is the
density of a topological observable that is related to $2$ and
$0$--observables by Witten's descent equation \wittop. The
currents in the bulk form the bulk algebra that is defined as
follows. Let us insert the $0$-observable $J_{V_2, (0)}$ at the
point $x$ in the bulk and integrate the current $J_{V_1}$ along a
small circle around $x$. This produces the zero observable at $x$,
which we will denote as $J_{[V_1,V_2]_{bulk},(0)}(x)$: \eqn\bulka{
\lim_{r \to 0} \la \int_{z \in S(x,r) } J_{V_1}(z) J_{V_2, (0)}(x)
\ra = \la  J_{[V_1, V_2]_{bulk}, (0)}(x) \ra}

\subsec{Generalities on 2d topological theories on surfaces with
boundaries} If a topological theory is defined on a surface with
boundaries, we must choose  boundary conditions that preserve
$Q_{BV}$. In particular, the current $J_{BV}$ that generates the
$Q_{BV}$ symmetry must vanish (within  correlation functions)  on
the boundary. Similarly, we have an additional condition for the
current $J_V$ to generate a symmetry: it must vanish when
restricted to the boundary within a correlation function:
\eqn\van{\la J_{V} \vert_{\p \Sigma} \ra = 0}

The boundary $0$-observables ${\CO}_{f}(x)$ are those that do not
change the vanishing of the current $J_{BV}$ when they are put on
the boundary $x \to {\p}{\Sigma}$. Here $f$ is some label on the
space of $0$-observables. In the type $C$ topological string $f$
stands for a function on the target space.

Since the energy-momentum tensor in topological theory is
$Q_{BV}$-exact, the correlators of local observables
 on the boundary are not changing when we
smoothly move the insertion point, without colliding with another
insertion point.

By moving the insertion points together we find that the
correlators
 of local observables on the
boundary are governed by the so call $*$-product on local boundary
observables: \eqn\stpro{ \langle\ldots \CO_{f_1}(x_1)
\CO_{f_2}(x_2)\rangle= \lim_{x_1 \rightarrow x_2} \langle\ldots
\CO_{f_1}(x_1) \CO_{f_2}(x_2)\rangle= \langle\ldots\CO_{f_1 *
f_2}(x_2)\rangle} if there are no other observables between
 $x_1$ and $x_2$. Here $x_1, x_2$ are two points on he boundary
 ${\p}{\Sigma}$.

Now, we can define the action
 $U_{V}$
of symmetry on the $0$-observable on the boundary $\CO_f(x)$ as
follows: \eqn\quantact{ \langle\ldots (\CO)_{U_V(f)}(x)  \rangle =
\lim_{r \rightarrow 0} \langle\ldots \int_{S(x,r)}J_{V}
\CO_{f}(x) \rangle   } where $\ldots$ stands for insertions of
other $Q_{BV}$-closed
 observables whose support (range of integration)
does not intersect the arc $S(x,r)$  of bulk points $y$ situated
at the distance $r$ from $x$. Notice that it contains quantum
corrections as compared to the naive classical symmetry action.

We will be interested in the computation of the Lie algebra of
actions on the boundary observables whose structure constants
$C_{ij}^{k}$ are defined as \eqn\defcijk{ U_{V_{i}}
(U_{V_{j}}(f))- U_{V_{j}} (U_{V_{i}}(f)) = C_{ij}^{k} U_{V_{k}}(f)
} As we will see below this algebra is different from the bulk
algebra defined in \bulka .

\subsec{Disk bubbling and deformation of the boundary algebra}

We are going to study here the correlator that determines the
commutator \defcijk\  in conformal topological theory :
\eqn\scom{\eqalign{& \la \CO_{[ U_{V_1} , U_{V_2} ](f)}(x) \rangle
= \la \int_{S(x,r_1)} J_{V_1} \int_{S(x,r_2)} J_{V_2} \CO_f(x)
\rangle - \cr & \la \int_{S(x,r_3)} J_{V_1} \int_{S(x,r_2)}
J_{V_2} \CO_f(x) \rangle \cr}} where $r_1 > r_2 > r_3$. Let us
denote the beginning of the $i$-th arc  and the end of the $i$-th
arc as $x_{-,i}$ and $x_{+,i}$ respectively.

It is clear that the r.h.s. of \scom\ is independent on the exact value
of $x_i$ - the only thing that matters is that the intervals
 $[x_{i,-} , x_{i,+}]$ are arranged as follows:
\eqn\intar{[x_{3,-} , x_{3,+}] \subset [x_{2,-} , x_{2,+}] \subset [x_{1,-} ,
 x_{1,+}]
}

\topinsert
\centerline{\epsfxsize=9.5cm\epsfbox{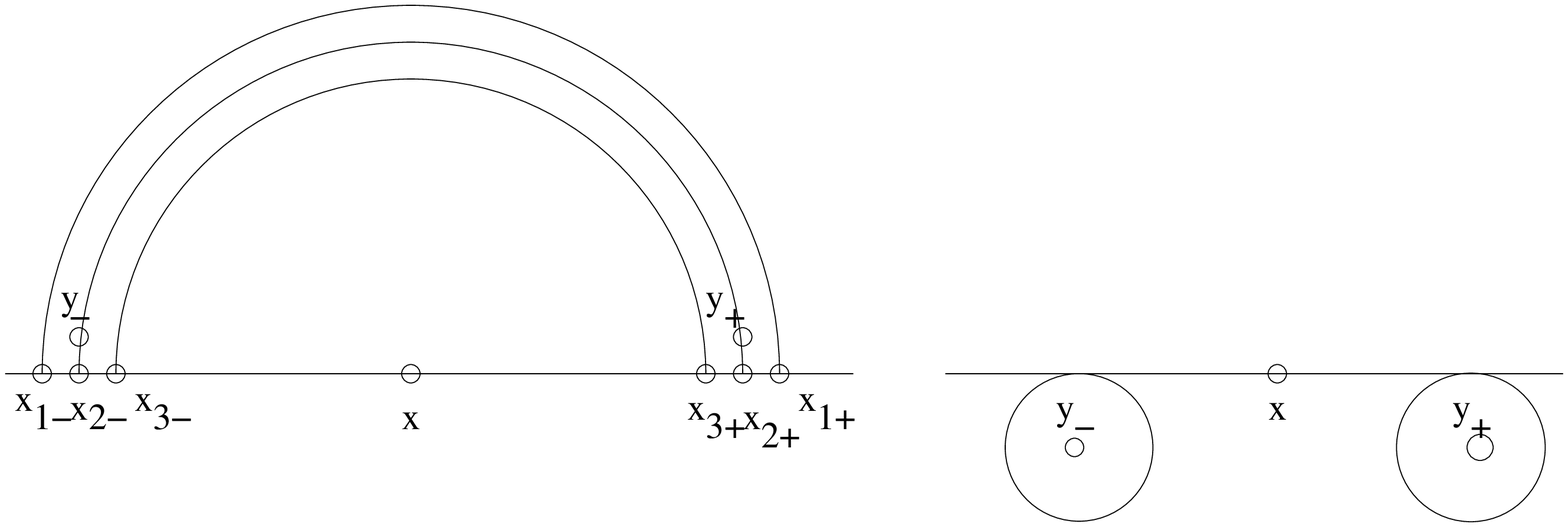}}

\endinsert

We will take the arcs close to each other, i.e. \eqn\ordr{ r_1 -
r_3 << r_3} The commutator appears as a result of deformation of
the arc $S(x,r_1)$ into the arc $S(x, r_3)$. The easiest way to
achieve this deformation would be to start with the replacement of
the arc $S(x,r_2)$ by the arc $S'(x, r_2)$ that connects the
points $y_{\pm}$ obtained from $x_{2,\pm}$ by very small shifts
inside the disk, namely \eqn\ordall{ |y_{\pm}-x_{2, \pm} | << r_1
- r_3 << r_3} Then, using the vanishing of current $J_{V_1}$ on
the boundary and applying Stokes theorem, we can rewrite the
commutator as a sum of two terms: \eqn\scom{\eqalign{ & \la\CO_{ [
U_{V_1} , U_{V_2} ](f)}(x)\rangle = \la \int_{y \in
\Gamma_{y_{-},y_{+}}} \int_{z \in \Gamma_{y}} J_{V_1}(z)
J_{V_2}(y) \CO_f(x) \rangle + \cr +  & \la \int_{y \in
\Gamma_{y_{-},y_{+}}} J_{V_2}(y) \int_{z \in SA} Q_{BV}(J_{V_1,
(2)}(z)) \CO_f(x) \rangle ,  } } where $\Gamma_{y_{-}, y_{+}}$ is
an arc that connects points $y_{-}$ and $y_{+}$, $\Gamma_{y}$ is a
small loop around point $y$, and $SA$ is the semi-annulus, whose
boundary consists of the two intervals $[x_{1,-}, x_{3,-}]$ ,
$[x_{1,+}, x_{3,+}]$ and the two arcs.

The first term in \scom\ is just the
 contribution from the OPE of the currents, i.e., is given by the bulk algebra
 \bulka .

The second term in \scom\ is more interesting. Taking $Q_{BV}$
from $J_{V_1,(2)}$ and applying it to $J_{V_2}$ we will get total
derivative $d J_{V_2,(0)}$ , see \bvqclo . We can integrate this
total derivative to get the following expression for the second
term: \eqn\secterm{ \eqalign{ & \la \int_{y \in
\Gamma_{y_{-},y_{+}}} J_{V_2}(y) \int_{z \in SA} Q_{BV}(J_{V_1,
(2)}(z)) \CO_f(x) \rangle = \cr & \la \int_{z \in SA} (
J_{V_2,(0)}(y_+)- J_{V_2,(0)}(y_-) ) J_{V_1, (2)}(z) \CO_f(x)
\rangle }} Now an interesting thing happens. In massive theory the
contribution of $SA$ would be negligible since its area is small.
In a conformal theory the notion of the absolute area makes no
sense. In order to see what really contributes, we make a
conformal transformation that maps the area around the points
$y_{\pm}$ to the disks with centers $y_{\pm}$ which are bubbled
out. The rest of the semi-annulus $SA$ is mapped into a figure
connecting these bubbled out disks - one can show that the
contribution of the rest of the $SA$ could be neglected.

The integral of $J_{V_1,(2)}$ over the bubbled disk $BD$, with the
operator  $J_{V_2, (0)}$ at its center, can be replaced by a
$0$-observable $\CO_{F_{BD}(V_1,V_2)}$ placed at the point where
bubbled disk joins the rest of the surface: \eqn\bddef{\la
\CO_{F_{BD}(V_1,V_2)} \ra = \la \int_{z \in BD} J_{V_2,(0)}(y)
J_{V_1, (2)}(z) \ra }

Thus,  we get the following expression for the commutator of the
$U_{V}$'s: \eqn\boundcomf{ [ U_{V_1} , U_{V_2} ](f) = U_{[V_1,
V_2]_{bulk}}(f)+ (f* F_{BD}(V_1,V_2) -F_{BD}(V_1,V_2)*f) } It is
the second term which only occurs due to the presence of the
boundary and it makes the algebra of boundary symmetries {\it
different} from that in the bulk.

\subsec{Example: $Diff$-symmetry of the $*$-product}

Now we will show how general considerations above about the action
of the symmetries on the boundary $0$-observables work in the $C$
model. The boundary $0$-observables in the model are the functions
on $X$ (one can also consider differential forms of higher degree,
but this does not give anything new), ${\CO}_{f} (z) = f (q(z))$.
The three-point function on the disk defines the $*$-product:
\eqn\strpr{\langle {\CO}_{f_1}(0) {\CO}_{f_2}(1) {\CO}_{f_3}
({\infty}) \rangle = \int_{X} {\n} \ f_1 * f_2 * f_3}where $\n$ is
some top degree form on $X$ (the descendant of BV measure $\n$).
It is convenient to work on the upper half-plane instead of the
disk and to replace $f_3 {\n}$ by a ${\d} (q({\infty}) - x)$, so
that \eqn\strp{\langle {\CO}_{f_1} (0) {\CO}_{f_2}(1) \rangle_{x}
= f_1 * f_2 (x) = f_1 f_2 (x) + {\pi}^{ij} {\p}_i f_1 {\p}_2 f_2 +
\ldots } where $\ldots$ for ${\Gamma} = {\tilde\Gamma} = 0$ are
given by the perturbation series, constructed in \maxim. If one
takes ${\pi} = const$, this series can be summed up, giving rise
to the so-called Moyal product: \eqn\moyal{f_1 * f_2 (x) = {\exp}
\left[ {\pi}^{ij} {{\p}\over{{\p \xi}^i}} {\p\over{\p\eta^j}}
\right] f_1 ({\xi}) g ({\eta}) \vert_{\xi = \eta = x}}

\subsec{ The bulk algebra of Poisson diffeomorphisms is classical}

Here we will not study the general symmetries of the $C$-model,
but we will restrict ourselves to the diffeomorphisms that
preserve the BV-action deformed by the Poisson bi-vector $\pi$,
i.e. \eqn\sympi{ \CL_{V} \pi = 0} Moreover we shall consider
Hamiltonian vector fields (on the simply-connected $X$ all Poisson
vector fields are Hamiltonian) : \eqn\poivfgh{ V_{h}^i = \pi^{ij}
\partial_{j} h(Q) }
One can check that such vector fields form the classical algebra:
\eqn\clasalg{ [ \CL_{V_{h}} , \CL_{V_{g}} ] = \{ h , g \}_{\pi} }
The currents that correspond to these classical diffeomorphisms
are the $1$-form components of the superfield $V_{h}^m (Q)
Q_{m}^{+}$ and are equal to \eqn\curre{ J_{V_{h}}=
V_{h}^{m}(q)p_{m}+ \p_k V_{h}^{m} p_m (0) q_{(1)}^k } It is not
trivial, one can check (Kontsevich technical Lemma \maxim) that
the bulk algebra of currents coincides with the classical algebra
\clasalg .

Now we are in position to show that the algebra of the action of
the currents on boundary observables is {\it not} classical, thus
presenting an example of how the  bubbling phenomenon works.

\subsec{Action of the currents on the boundary observables in the
$C$ model}

Consider the boundary observable $\CO_f(x)$ that corresponds  to
the function  $f(q)$ placed  at  point $x$. The current $J_{V_h}$
is integrated along the arc with the endpoints, that we will
denote as $x_{-}$ and $x_{+}$.

If we consider $\pi$ perturbatively, the $U_{V_h}$ operation would
be a series in powers of bi-vector $\pi$: \eqn\uvop{U_{V_h} (f)
(x) = \langle {\CO}_{f} (0) \int_{x_{-}}^{x_{+}}  J_{V_{h}}
\rangle_{x}} The leading term is equal to the classical expression
$\CL_{V} f$, but there are other terms. One can explicitly compute
them, using Konstevich diagrams, but instead we can use a
shortcut.

We use the fact  that the $1$-observable that corresponds to the
Poisson vector field is a sum of a $Q_{BV}$-exact term and a total
derivative, namely: \eqn\trick{ (V_{ h}^m Q_{m}^{+})_{(1)}= dh +
Q_{BV}(h) } Thus, within correlator one has: \eqn\trickcor{
\int_{S(x,r)} (V_{h}^m Q_{m}^{+})_{(1)} = \int_{S(x,r)} dh =
\lim_{y \rightarrow x_{+}}h - \lim_{y \rightarrow x_{-}}h } Now
let us assume that ${\pi} = const$. One can show that in this case
the limit coincides with the boundary value \cf, and we get that
\eqn\tostar{ \la \int_{S(x,r)}J_{V_{ h}} \CO_f(x)\ra = \la
\CO_f(x) \CO_h(x_+)\ra - \la \CO_f(x) \CO_h(x_-)\ra} The r.h.s. of
\tostar\ can be computed with the help of the star-product.

Thus, from equation \tostar, we obtain the following result for
$U_{V_{h}}$: \eqn\starc{U_{V_{h}}(f)= h*f - f*h= [ h ,^{\kern
-.05in *} f ] } An obvious calculation shows that  $U_V$ commute
via $*$-commutator rather than via Poisson commutator. The
difference is just the manifestation of the disk bubbling
phenomena, mentioned above.

In the case ${\pi} = const$ one can actually use this trick and
compute all Kontsevich diagrams for an arbitrary vector field $V$,
not necessarily Poisson. One finds: \eqn\uvops{U_{V} f (x) = {\hat
A}^{-1} \left[ {\pi}^{ij} {{\p}\over{{\p \xi}^i}}
{{\p}\over{{\p}{\eta}^j}} \right] V^i ({\xi}) {\p}_i f ({\eta})
\vert_{\xi = \eta = x}} where ${\hat A}^{-1} (z) = {{e^{z} -
e^{-z}}\over{z}}$. This operation on functions has already
appeared in \stress.

\newsec{Discussion and conclusions}

In this paper, we have considered the realization of the target
space diffeomorphisms in the type $C$ topological string.
Specifically we looked at the symmetries of closed and open string
theories. In particular, we analyzed the algebra of charges which
generate
 the infinitesimal transformations preserving closed
string background ${\pi}$. The charges are given by the integrals
of currents over little circles surrounding the bulk observables,
and little arcs surrounding the boundary observables.
 We found
that the charges, which generate infinitesimal transformations of
the boundary observables,  form an algebra, that is a deformation
of the algebra of bulk charges, by the contribution of the bubbled
disks.

The worldsheet perturbation technique developed in \maxim\ can be
applied to define infinitesimal transformations $U_{V}$ in more
general context,  corresponding to general, not necessary Poisson
vector fields $V$: \eqn\genk{U_V f = V^i {\p}_i f + {\pi}^{ij}
{\p}_i V^m {\p}^2_{jm} f + \ldots .} They are a specific component
of the $L_{\infty}$-morphism of Kontsevich: $$ U_V f =
\sum_{n=0}^{\infty} {1\over{n!}} {\CF}_1 ( V,{\pi}, \ldots
{\scriptstyle n \ times} \ldots, \pi ) [f] $$One can also
establish that these operations make the $*$-product covariant in
the sense that \eqn\covstr{U_V ( f * g) - (U_V f) * g - f * (U_V
g) = {\CL}_V {\pi} {{\d}\over{\d \pi}} ( f * g)} and that they
form the algebra \eqn\dfal{[U_{V_1}, U_{V_2}] f - U_{[V_1, V_2]} f
= [ F_{BD}(V_1 , V_2) ,^{\kern -.05in *} f]} where $F_{BD}(V_1,
V_2) = \sum_{n=1}^{\infty} {1\over{n!}}{\CF}_{0} (V_1, V_2, {\pi},
\ldots {\scriptstyle n \ times}, \ldots ) $.

These properties can be formally established from the
$L_{\infty}$-morphism properties \old\manchon\cfnew, but it is
instructive to understand them from the world-sheet point of view.
We showed that the deformation of the algebra is due to the
phenomenon of disk ``bubbling'' which is purely field-theoretic
effect.

This result can be applied to construct a ``covariant'' action of
a \nc\ scalar field theory on ${\IR}^{n}_{\pi}$ - the
noncommutative space, whose algebra of functions is the
$*$-product algebra. The field of this theory is an element
${\phi}$ of the algebra ${\IR}^n_{\pi}$, and the action is given
by: \eqn\covac{{\CS} = \int_{{\IR}^n} g^{ij} [ D_i, {\phi} ]* [
D_j, {\phi} ] + V ({\phi}) } where $V ({\phi})$ is some polynomial
function, say ${\half} m^2 {\phi}^2 + {\l \over {4!}} {\phi} *
{\phi} * {\phi} * {\phi}$, and $g^{ij}$ is a constant matrix. Then
\covstr\ implies that the correlation functions in such theory
will be invariant under the transformations:
\eqn\covtrns{\eqalign{ & D_i \mapsto D_i + U_V D_i \cr & {\pi}
\mapsto {\pi} + {\CL}_V {\pi} \cr & {\phi} \mapsto {\phi} + U_{V}
{\phi} \cr}} which could be used to define an improved
stress-energy tensor (cf. \stress).

\footatend\vfill\supereject\immediate\closeout\rfile\writestoppt
\baselineskip=14pt\centerline{{\bf References}}\bigskip{\frenchspacing%
\parindent=20pt\escapechar=` \input refs.tmp\vfill\eject}\nonfrenchspacing
\bye